\documentclass{article}
\usepackage[T1]{fontenc}
\usepackage[utf8]{inputenc}
\usepackage{ismir} 
\usepackage{amsmath, amssymb, cite,url}
\usepackage{graphicx}
\usepackage{color}
\usepackage{appendix}
\usepackage{arydshln}
\title{Interpretable and Perceptually-Aligned Music Similarity with Pretrained Embeddings}

\twoauthors
  {Arhan Vohra} {Universitat Pompeu Fabra \\ Barcelona, Spain}
  {Taketo Akama} {Sony Computer Science Laboratories \\ Tokyo, Japan}

\begin{document}

\maketitle

\begin{abstract}
Perceptual similarity representations enable music retrieval systems to determine which songs sound most similar to listeners. State-of-the-art approaches based on task-specific training via self-supervised metric learning show promising alignment with human judgment, but are difficult to interpret or generalize due to limited dataset availability. We show that pretrained text-audio embeddings (CLAP and MuQ-MuLan) offer comparable perceptual alignment on similarity tasks without any additional fine-tuning. To surpass this baseline, we introduce a novel method to perceptually align pretrained embeddings with source separation and linear optimization on ABX preference data from listening tests. Our model provides interpretable and controllable instrument-wise weights, allowing music producers to retrieve stem-level loops and samples based on mixed reference songs.
\end{abstract}

\section{Introduction}

Music retrieval and recommendation systems exploit several notions of music similarity; these can be represented through \textit{contextual} cues (e.g. coincident listening behavior and geographical proximity of artists) or the closeness of \textit{content-based} features (tempo, MFCCs, instrumentation etc.) \cite{zeng_survey_2024}. While such approaches have proven highly effective for scaling music discovery, they provide only implicit approximations of \textit{perceptual similarity}, the extent to which two pieces \textit{sound} similar to listeners.

Consider a music producer searching for the perfect snare drum sound within a library of thousands of samples. Traditional keyword-based search requires samples to be meticulously tagged and is bottlenecked by the user's ability to \textit{describe} their target sound. Content-based retrieval methods such as Query-by-Example (QbE) offer a more intuitive alternative: users can provide an audio example to retrieve similar sounds without needing to articulate specific timbral qualities. However, the effectiveness of QbE systems hinges on the underlying similarity model's ability to capture how human listeners actually perceive and compare sounds.

These limitations become even more pronounced when searching within mixed musical recordings. A producer might hear a snare sound in a complete track and wish to find similar drum hits in their sample library, or identify songs that feature comparable instrumental textures. Such queries require not only perceptual alignment but also the ability to disentangle instrument-specific similarities from the complex acoustic mixture. If a model can learn to isolate and compare individual instrumental components within mixed tracks while maintaining perceptual validity, it could produce transformative workflows, such as retrieving similar drum patterns from just a reference song, or finding bass lines with a specific timbral character in one shot. 

Recent work in music similarity has sought to achieve perceptually-aligned outcomes through metric learning models fine-tuned on data from large-scale listening studies \cite{hashizume_investigation_2025, hashizume_music_2022, hashizume_learning_2024, imamura_music_2025}. These methods demonstrated strong agreement with human judgments, but face notable limitations: they demand task-specific training, are often trained and tested on closely related datasets, and offer limited interpretability.

Simultaneously, general-purpose pretrained audio embeddings such as CLAP, MERT, and MuQ-MuLan \cite{wu_large-scale_2023, li_mert_2024, huang_mulan_2022, zhu_muq_2025} have eclipsed handcrafted features and task-specific models at various tasks by leveraging text-audio alignment to create semantically rich latent spaces. While CLAP embeddings were observed to excel at semantic retrieval \cite{araz_evaluation_2024} and timbre similarity alignment \cite{tian_assessing_2025}, their performance has not been evaluated against any of the state-of-the-art music similarity representations.

In this study, we address these gaps in two ways:

\begin{enumerate}

    \item \textbf{We evaluate CLAP and MuQ embeddings directly against trained similarity representations} on perceptual benchmarks, using the Inst-Sim-ABX dataset of human similarity judgments \cite{wu_large-scale_2023, zhu_muq_2025, hashizume_investigation_2025}. 

    \item \textbf{We propose an explicit and interpretable instrument-aware similarity model}, which disentangles perceptual similarity by computing a weighted sum of instrument-wise similarities based on pretrained embeddings. This approach leverages music source separation (MSS) to isolate instrument stems, allowing us to examine how different components of a song contribute to human perceptions of similarity.

\end{enumerate}
\section{Related Work}

\subsection{Psychoacoustic Models of Music Similarity}
Human perception of music similarity is not a static function of acoustic distance but a dynamic cognitive process. Listeners process auditory cues by actively filtering them through an internal "world model" informed by contextual cues, and ultimately modulated by their attention \cite{herre_how_2003}. This process is governed by global precedence, suggesting that listeners identify global structures such as melody contour, texture, and tempo, faster and more dominantly than local details \cite{sanders_local_2007}. Neuroscientific studies have shown that musicians employ selective attention mechanisms to isolate specific melodies in a mixture \cite{manting_how_2025}. Consequently, effective similarity models for production tasks must be capable of replicating perceptual disentanglement processes either implicitly or explicitly.

\subsection{Learned Representations for Music Similarity}
Early computational approaches modeled songs as "bags-of-frames", representing audio as distributions of spectral features compared via Earth Mover’s Distance or KL divergence \cite{logan_music_2001, Aucouturier2002MusicSM}. While this established a robust baseline for timbre comparison, it discarded temporal structure and suffered from the "hubness" phenomenon, where generic tracks dominated retrieval results \cite{wang_2011, knees_survey_2013}. Subsequent work sought to mitigate these limitations by augmenting spectral models with handcrafted features designed to capture temporal and perceptual attributes, such as fluctuation patterns and rhythm histograms \cite{herre_how_2003, DBLP:conf/ismir/PampalkFW05, pohle_rhythm_2009, bogdanov_unifying_2011}. 

To better align these feature sets with semantics, early metric learning approaches were adopted to optimize similarity functions using contextual data, such as artist tags or user listening history \cite{DBLP:conf/ismir/SlaneyWW08, west_model-based_2006}. Later, deep metric learning (DML) shifted the paradigm from feature engineering to representation learning. Lee et al. showed Conditional Similarity Networks (CSNs) with triplet loss to be particularly useful in learning a disentangled embedding space, where specific dimensions are activated via binary masks to represent distinct notions of similarity (timbre, genre, mood etc.) \cite{lee_disentangled_2020}.

Building upon this, Hashizume et al. proposed several DML approaches to disentangle instrumental similarity subspaces. The "Cascade" model uses automatic source separation to split mixed songs into instrumental stems, each passed to a dedicated CNN trained with triplet loss \cite{hashizume_music_2022, imamura_music_2025}. Meanwhile the "disentangled CSN" (D-CSN) model learns to implicitly separate mixed sounds by training instrumental subspaces through masking \cite{hashizume_learning_2024}. Both of these models rely on the S4 assumption: that different randomized segments of the same song sound similar to one another, which is used to create training data for self-supervision via triplet loss. To test the perceptual alignment of music similarity models, the Inst-Sim-ABX dataset was introduced, and later used to fine-tune the Cascade and D-CSN models with human preferences \cite{hashizume_investigation_2025, imamura_music_2025}. These task-specific models currently represent the state-of-the-art in perceptual alignment but require specialized, resource-intensive training pipelines.

\subsection{Foundation Text-Audio Models}
General-purpose foundation models have emerged as powerful alternatives to task-specific machine learning approaches. Contrastive Language-Audio Pretraining (CLAP) aligns audio encodings with text descriptions, creating a joint multimodal space where similarity is defined by semantic proximity. MuQ-MuLan extends this approach to be musically-aware by focusing on intrinsic acoustic structure through masked modeling, using a Mel-Residual Vector Quantization (Mel-RVQ) tokenizer to capture fine-grained spectro-temporal details. While these models excel at downstream tasks like zero-shot classification and tagging, their alignment with human perceptual similarity remains underexplored.
\section{Methodology}
\subsection{Dataset and Perceptual Similarity Annotations}

We adopt the Inst-Sim-ABX dataset introduced by Hashizume and Toda \cite{hashizume_investigation_2025}, which provides perceptual similarity annotations for audio segments extracted from the Slakh corpus. The original Slakh2100 corpus \cite{manilow_cutting_2019} was designed to study music source separation and thus contains multitrack stems and full-mix audio for a large number of synthesized songs. The Inst-Sim-ABX dataset contains annotations for 5-second audio segments derived from a subset of 136 tracks in Slakh, prepared as either individual instrumental stems (bass, drums, piano, guitar, residuals) or full-mix audio. Segments were organized into ABX triplets, where a reference segment X is paired with two comparison segments for subjects to judge. Importantly, the segments representing each instrument class were sampled separately from Slakh. As a result, the ABX triplets for drums, bass, guitar, etc. are independent datasets rather than parallel views of the same musical material, and thus cannot be compared across instruments on a one-to-one basis.

A large-scale perceptual listening test was conducted with 281 unique subjects. In each trial, participants judged which comparison was perceptually more similar to the reference. Two test configurations were defined:

\begin{itemize}
    \item \textbf{XAB:} X, A, B are each extracted from three different tracks in the Slakh dataset
    \item \textbf{XYC:} The reference segment X and the comparison segment Y are taken from the same track, while the last segment C is from a different track
\end{itemize}

For each triplet, responses were aggregated via majority vote. Following prior work, we retained only those triplets with high inter-subject agreement in order to ensure reliable perceptual alignment. We additionally adopted this filtering to maintain comparability with earlier studies, which reported results under different confidence thresholds. Accordingly, we present results at multiple cutoff levels (e.g., 75\% and 80\%) to align with the evaluation settings used in previous work.

\subsection{Audio Preprocessing and Stem Extraction
}
For our study, we primarily used the audio segments specified in the Inst-Sim-ABX dataset, resampled from 44.1kHz to 48kHz for the CLAP audio encoder. We extended the full-mix subset of this data for our weighted similarity model by extracting two types of instrumental stems for each mixed song segment $S_{\text{full-mix}}$.

\subsubsection{Automatic MSS Stems (Demucs)}
We applied the open-source Demucs v4 music source separation model \cite{rouard_hybrid_2022}. We utilized two configurations: the standard htdemucs model, which separates sources into four categories (vocals, drums, bass, other), and the htdemucs\_6s model, which additionally extracts guitar and piano stems, allowing for higher-granularity analysis. Since the Slakh dataset contains no vocals, the predicted vocal stems are effectively silent. The remaining stems provide approximate instrumental decompositions that allow us to embed and compare instrument-wise representations even for the full-mix segments of Inst-Sim-ABX.

\subsubsection{Ground-truth Stems (Slakh)}
As Slakh is derived from MIDI-driven virtual instrument renderings, it provides access to perfectly isolated multitrack audio for each song. We therefore also extracted true stems directly from the Slakh tracks, focusing on the same fundamental categories used in Demucs: bass, drums, guitar, piano and residuals (all remaining instruments). These ground-truth stems serve as an upper bound for instrument separation quality, as they contain no bleed or separation artifacts.

\subsection{Pretrained Embeddings}
To represent audio segments in a common latent space, we compared two audio encoders: \textbf{LAION-CLAP} (checkpoint: 630k-audioset-best.pt, which is recommended for music tasks) and \textbf{MuQ-MuLan} (checkpoint: MuQ-MuLan-large). Both approaches use contrastive learning to align text-audio pairs, with the same text encoder (RoBERTa) \cite{DBLP:journals/corr/abs-1907-11692} albeit different audio encoders and training datasets.

LAION-CLAP maps audio and text into a shared 512-dimensional space optimized for semantic alignment. Its audio encoder uses a HTS-AT (Hierarchical Token-Semantic Audio Transformer) architecture, which processes audio as a hierarchy of local and global features to handle variable-length inputs efficiently \cite{DBLP:journals/corr/abs-2202-00874}. The specific checkpoint used in this study was trained on LAION-Audio-630k, a massive dataset containing over 630,000 audio-text pairs (>4,300 hours) covering diverse domains, including music, speech, and environmental sounds. While the model is designed for a 10-second input window, it natively handles shorter inputs (like our 5-second segments) by repeating and padding the signal to fill the required duration.

MuQ-MuLan is a specialized music-text model based on the MuQ audio encoder, which is optimized for musical understanding tasks \cite{zhu_muq_2025}. MuQ employs a Conformer (convolution-augmented transformer) backbone pre-trained via self-supervised learning to predict discrete acoustic tokens from a Mel Residual Vector Quantization (Mel-RVQ) tokenizer. This objective allows the model to capture fine-grained musical structures, such as pitch and harmony, before being aligned with text. We utilized the open-source release of MuQ-MuLan, which is trained on 900 hours of musical audio from the Music4all dataset \cite{9145170}.

\subsection{Similarity Models}
To predict perceptual similarity, we evaluate two distinct modeling approaches that operate on the frozen embeddings produced by the previously described pretrained encoders.

Let $\Phi(\cdot)$ denote the embedding function corresponding to a chosen encoder (i.e., either LAION-CLAP or MuQ-MuLan). For any given audio segment $x$, $\Phi(x) \in \mathbb{R}^{512}$ represents its $512$-dimensional feature vector in the shared latent space.

\subsubsection{Standard Cosine Similarity}
Our first model adopts the assumption that perceptual similarity is equivalent to the cosine similarity between global embeddings in the latent space. 

For an XAB triplet consisting of a reference $X$ and candidates $A$ and $B$, the model predicts the more similar candidate by maximizing the cosine score:

\begin{equation}
    \hat{y} = \operatorname*{argmax}_{V \in \{A, B\}} \left( \frac{\Phi(X) \cdot \Phi(V)}{\|\Phi(X)\| \|\Phi(V)\|} \right)
\end{equation}

\subsubsection{Instrument-wise Weighted Cosine Similarity}
We propose a second, explicit similarity model which posits that the perceptual similarity of a full mix is a weighted sum of the similarities of its constituent instrumental parts. 

This approach leverages Music Source Separation (MSS) to decompose the mixed audio into discrete instrument stems (e.g., bass, drums, vocals, other). We define the similarity feature vector $\mathbf{f}$ for a given triplet as the difference in component-wise similarities between candidate A and candidate B. Specifically, for each instrument class $k \in \{\text{bass}, \text{drums}, \dots, \text{mix}\}$:

\begin{equation}
    f^{(k)} = \cos(\Phi(X^{(k)}), \Phi(A^{(k)})) - \cos(\Phi(X^{(k)}), \Phi(B^{(k)}))
\end{equation}

where $X^{(k)}$ denotes the $k$-th stem of the reference track. The triplet is then represented by the vector $\mathbf{f} = [f^{\text{bass}}, f^{\text{drums}}, \dots, f^{\text{mix}}]$.

To learn the perceptual importance of each instrument, we construct a design matrix $\mathbf{F} \in \mathbb{R}^{N \times K}$ by stacking the feature vectors of all $N$ triplets in the dataset. We then model the human judgment $y \in \{-1, 1\}$ (where 1 indicates preference for A) using linear regression:

\begin{equation}
    \hat{y} = \mathbf{w}^T \mathbf{f}
\end{equation}

We enforce a zero intercept to preserve the symmetry of the task: if candidates A and B are identical (resulting in $\mathbf{f}=\mathbf{0}$), the predicted preference should be neutral ($\hat{y}=0$). We report learned weights $\mathbf{w}$ using both Ordinary Least Squares (OLS) and Ridge Regression. Additionally, we compare performance using stems separated automatically by Demucs as well as ground-truth stems from Slakh. This allows us to disentangle limitations caused by separation artifacts from the inherent validity of the weighted similarity hypothesis.

Additionally, we evaluated our approach using both 4-source and 6-source music source separation models from Demucs. The choice of configuration has important trade-offs. The 4-source model tends to be more robust, since its broad categories (bass, drums, other, mix) are consistently present across tracks and less prone to separation errors. By contrast, the 6-source model introduces dedicated stems for guitar and piano, but these tracks are more vulnerable to bleeding and artifacts \cite{rouard_hybrid_2022}, reducing their reliability. Still, using a larger number of sources should in principle enable richer perceptual modelling: collapsing everything except bass and drums into a single “residuals” category risks losing much of the melodic and harmonic structure that listeners rely on for similarity judgments.

\subsection{Evaluation Process}
For both similarity models, the primary metric is \textbf{perceptual agreement}: the proportion of triplets where the model's predicted preference aligns with the majority choice of human listeners. This metric is consistent previous studies that used the Inst-Sim-ABX dataset, giving us clear baselines for our results \cite{hashizume_investigation_2025, imamura_music_2025}. As the previous studies were closed-source, we use their reported results as our baseline data.

Given the limited size of the Inst-Sim-ABX dataset, we also employed a cross-validation strategy to understand the variance in our regression results and avoid overfitting. We performed 100 iterations of stratified shuffle-split cross-validation, where each iteration randomly partitioned the dataset into 70\% training and 30\% testing sets while preserving the proportion of class labels. The reported accuracy and standard deviation values are aggregated over these 100 distinct splits.

\section{Results}

\subsection{Standard Cosine Similarity}
We first evaluate the baseline performance of pretrained embeddings by computing the standard cosine similarity between full-mix audio segments.

\textbf{Table \ref{tab:cosine_results}} presents the perceptual agreement accuracy for both CLAP and MuQ-MuLan alongside state-of-the-art task-specific models from prior literature. Both foundation models demonstrate strong transfer capabilities without any fine-tuning. In the XAB configuration (different tracks), MuQ-MuLan achieves an accuracy of 72.4\% on fully mixed songs, slightly outperforming CLAP (71.9\%) and remaining competitive with the supervised Cascade-PAFT model. Notably, MuQ-MuLan excels in the XYC configuration (same-track segments), achieving near-perfect agreement for drums (97.1\%) and full-mix (96.6\%). This suggests that the MuQ encoder, with its specific pre-training on musical structures and quantization targets, is perhaps highly sensitive to the rhythmic and harmonic signatures that define a track's identity.

However, the generic embedding models struggle with fine-grained similarity in some cases. For bass and guitar similarity, the task-specific Cascade-PAFT model significantly outperforms both CLAP and MuQ-MuLan (e.g., 75.8\% for guitar vs. $\approx$66\% for the foundation models). Perhaps Cascade-PAFT's instrument-specific CNN outputs can sometimes capture richer instrument-aware similarity representations, but unlike CLAP and MuQ-MuLan, this model was trained and fine-tuned on audio examples drawn from the same underlying distribution as the perceptual testing stimuli (Slakh) \cite{imamura_music_2025}. As such, the generalizability of Cascade-PAFT cannot be verified without additional perceptual testing based on unseen datasets.

\begin{table*}[t]
\centering
\begin{tabular}{lcccccc}
\hline
\textbf{Model} & \textbf{Drums} & \textbf{Bass} & \textbf{Guitar} & \textbf{Piano} & \textbf{Residuals} & \textbf{Mix} \\
\hline
\multicolumn{7}{c}{\textit{XAB (all-different)}} \\
CLAP-cosine & 78.5 & 69.8 & 64.2 & \textbf{68.5} & 64.3 & 71.9 \\
MuQ-MuLan-cosine & \textbf{79.3} & 71.6 & 66.7 & 66.3 & \textbf{65.3} & \textbf{72.4} \\
\hdashline
D-CSN (Hashizume et al.) & 52.7 & 67.5 & 60.1 & 58.5 & 58.3 & -- \\
Cascade-PAFT (Imamura et al.) & 70.5 & \textbf{76.3} & \textbf{75.8} & 64.6 & -- & -- \\
\hline
\multicolumn{7}{c}{\textit{XYC (one-shared)}} \\
CLAP-cosine & 94.8 & 92.3 & 87.2 & 94.3 & 91.4 & 89.3 \\
MuQ-MuLan-cosine & \textbf{97.1} & \textbf{92.4} & 87.0 & \textbf{95.2} & 87.7 & \textbf{96.6} \\
\hdashline
D-CSN (Hashizume et al.) & 94.8 & 88.6 & \textbf{92.0} & 87.1 & \textbf{91.6} & -- \\
Cascade-PAFT (Imamura et al.) & 95.9 & 90.1 & 91.2 & 93.1 & -- & -- \\
\hline
\end{tabular}
\caption{Perceptual similarity agreement (\%) for cosine similarity models compared with state-of-the-art baselines. Results for D-CSN and Cascade-PAFT are reported from \cite{hashizume_investigation_2025} and \cite{imamura_music_2025} respectively. The best performance for each column is highlighted in bold.}
\label{tab:cosine_results}
\end{table*}

\subsection{Instrument-wise Weighted Cosine Similarity} 

We evaluate our proposed method of computing similarity as a weighted sum of separated instrument stems.

\textbf{Table \ref{tab:clap_weighted_results}} summarizes the performance for CLAP. We observe that incorporating instrument stems consistently degrades performance relative to the full-mix baseline. Using automatically separated stems (Demucs) in the 6-stem configuration drops accuracy from 84.6\% to 83.2\%. Even with perfect ground-truth stems from Slakh, CLAP yields only marginal gains ($\sim$1\%).

In contrast, MuQ-MuLan benefits significantly from the weighted similarity approach, as shown in \textbf{Table \ref{tab:muq_weighted_results}}. While the 4-stem model (which groups melodic instruments into "residuals") offers no improvement over the baseline, the 6-stem configuration achieves a substantial boost. Using Demucs stems, the ridge regression model reaches an accuracy of 90.4\%, the highest overall result observed in our study and a notable improvement over the 86.8\% baseline.

Crucially, this improvement is robust to source separation artifacts; the model performs even better with Demucs stems (90.4\%) than with ground-truth stems (87.4\%). This counter-intuitive result may imply that separation artifacts preserve contextual cues that aid the MuQ encoder, whereas perfectly dry ground-truth stems might lie further out of the distribution of MuQ's pre-training data. \textbf{Table \ref{tab:all_model_weights}} shows the average instrument weights assigned in our experiments. Guitar stems (which are mixed into "residuals" in the 4-stem models) are highly weighted in the 6-stem configurations. This suggests that extracting individual stems may produce some perceptually relevant form of information gain that is otherwise lost in the global mix embedding.

\begin{table}[t]
\centering
\begin{tabular}{l c c}
\hline
\textbf{Configuration}& \textbf{OLS} & \textbf{Ridge} \\
 \multicolumn{3}{c}{4 Stems + Mix}\\
MSS stems (Demucs)       & $84.5 \pm 3.0$ & $84.1 \pm 2.9$ \\
Ground-truth stems (Slakh) & $85.9 \pm 3.1$ & $85.7 \pm 3.1$ \\ \\
 \multicolumn{3}{c}{6 Stems + Mix}\\
 MSS stems (Demucs)& 83.6 $\pm$ 3.5&83.2 $\pm$ 3.3\\
 Ground-truth stems (Slakh)& 85.4 $\pm$ 3.4&84.6 $\pm$ 3.4\\
\hline
\textbf{Baseline (CLAP-cosine)} & \multicolumn{2}{c}{$84.6 \pm 0.0$} \\
\hline
\end{tabular}
\caption{Instrument-wise weighted CLAP model: perceptual similarity agreement (mean \% $\pm$ std) across 100 stratified 70/30 training splits. Results include automatic separation (MSS) and true stems from Slakh.}
\label{tab:clap_weighted_results}
\end{table}

\begin{table}[t]
\centering
\begin{tabular}{l c c}
\hline
\textbf{Configuration}& \textbf{OLS} & \textbf{Ridge} \\
 \multicolumn{3}{c}{4 Stems + Mix}\\
MSS stems (Demucs)        & $87.0 \pm 3.3$ & $87.0 \pm 3.3$ \\
Ground-truth stems (Slakh) & $87.6 \pm 3.1$ & $87.1 \pm 3.1$ \\ \\
 \multicolumn{3}{c}{6 Stems + Mix}\\
 MSS stems (Demucs)& \textbf{90.1 $\pm$ 2.4}&\textbf{90.4 $\pm$ 2.3}\\
 Ground-truth stems (Slakh)& 87.3 $\pm$ 2.9&87.4 $\pm$ 2.7\\
\hline
\textbf{Baseline (MuQ-cosine)} & \multicolumn{2}{c}{$86.8 \pm 2.7$} \\
\hline
\end{tabular}
\caption{Instrument-wise weighted MuQ-MuLan model: perceptual similarity agreement (mean \% $\pm$ std) across 100 stratified 70/30 training splits. Results include automatic separation (MSS) and true stems from Slakh. }
\label{tab:muq_weighted_results}
\end{table}

\begin{table*}[t]
\centering
\begin{tabular}{l ccccccc}
\hline
\textbf{Model Configuration} & \textbf{Bass} & \textbf{Drums} & \textbf{Guitar} & \textbf{Piano} & \textbf{Vocals} & \textbf{Residuals} & \textbf{Mix} \\
\hline
\multicolumn{8}{l}{\textit{CLAP}} \\
4-stem & 0.38 & 0.37 & — & — & 0.10 & 1.42 & 1.19 \\
6-stem & 0.11 & 0.41 & 0.43 & 0.10 & 0.21 & 0.57 & 1.92 \\
\hline
\multicolumn{8}{l}{\textit{MuQ-MuLan}} \\
4-stem & 0.25 & 0.78 & — & — & 0.08 & 0.86 & 1.94 \\
6-stem & 0.32 & 0.83 & 0.61 & 0.27 & 0.33 & 1.27 & 1.64 \\
\hline
\end{tabular}
\caption{Learned instrument weights ($\mathbf{w}$) for all weighted similarity configurations. The 4-stem model does not extract guitar and piano stems, instead relegating them to "residuals" if present.}
\label{tab:all_model_weights}
\end{table*}

\section{Discussion}

Our results highlight the efficacy of semantic audio embeddings for perceptual music similarity, even when applied zero-shot. Both CLAP and MuQ-MuLan embeddings, without any task-specific fine-tuning, achieve performance comparable to specialized models like D-CSN \cite{hashizume_investigation_2025}, demonstrating that large-scale pretraining on diverse audio-text pairs captures significant perceptually relevant information.

We hypothesize that this alignment arises because the contrastive learning objective forces the model to adopt a feature weighting mechanism analogous to human auditory perception. To successfully map audio to natural language descriptions (for instance, distinguishing 'breakbeat' from 'reggaeton') the model cannot rely on raw spectral matching; it must instead implicitly learn to attend to the specific acoustic cues (such as rhythmic syncopation or timbral texture) that humans prioritize when categorizing sound. Consequently, the resulting embeddings are not merely acoustic representations but are effectively shaped by the same cultural and semantic frameworks that structure human similarity judgments. By learning how to label, these models also learn how to \textit{listen}.

To extend our baseline findings, we introduced an instrument-wise weighted similarity model which disentangles mixed songs into instrumental stem embeddings. We showed that the MuQ-based version of this similarity model produces state-of-the-art perceptual results on the Inst-Sim-ABX dataset. This demonstrates that we can improve the perceptual alignment of pretrained embeddings by integrating psychoacoustic domain knowledge without the need for expensive fine-tuning. Furthermore, unlike black-box metric learning approaches, our method provides an interpretable mechanism for disentangling perceptual relevance, allowing us to quantitatively verify whether human similarity judgments are driven by specific instrumental cues. This explicit parameterization opens new doors for user-centric retrieval interfaces, where listeners or producers could intuitively navigate large music libraries by manually adjusting importance sliders for drum-similarity or guitar-similarity between songs.

Our results must be interpreted within the constraints of the Inst-Sim-ABX dataset. First, the source material derived from Slakh2100 consists entirely of MIDI-synthesized audio. Consequently, our evaluation cannot account for the complexities of real acoustic recordings, vocal nuances, or the varying production qualities found in commercial music. Second, the dataset size is relatively small; after filtering for high inter-subject agreement, we retained only $\approx$330 triplets. This limited sample size raises the risk that our results may not fully generalize to larger, more diverse musical collections. 

Furthermore, a critical analysis of the learned regression weights suggests that the improvements in the weighted similarity model may be partially idiosyncratic to this dataset. To contextualize this within the 6-stem separation framework: by explicitly removing drums, bass etc., the 'residuals' channel retains all remaining harmonic and textural content, including synthesizers, strings, and brass. Given that the Slakh dataset relies heavily on synthesized MIDI arrangements, this stem likely contains the defining timbral 'texture' of many tracks. So the reliance on this stem may reflect a sensitivity to these musical components due to the dataset's distribution. Future work using acoustic datasets may be needed to disentangle this effect further.

This suggests that while the model is effective, the learned weights may not yet represent stable, universally applicable preferences, but rather a fit to the specific synthetic style of the Slakh corpus. These preferences are also unlikely to be constant across musical traditions - for example, we may attend to timbral and rhythmic cues differently in Western pop music compared to Indian classical music. Future work using real-world multi-track recordings could help verify if these findings hold up beyond the synthetic domain.

\bibliography{references}

\end{document}